%% file: main.tex
\documentclass[letterpaper, 10 pt, conference]{ieeeconf}  % Comment this line out
\usepackage[pdftex,pdfauthor={Quindlen},pdftitle={Active Learning for Data-Driven Verification of Dynamical Systems}]{hyperref}
\hypersetup{colorlinks,linkcolor={green!50!black},citecolor={green!50!black},urlcolor={blue!80!black}}
\makeatletter \let\NAT@parse\undefined \makeatother

\input{jfq_includes}

\usepackage{theorem,balance}
\usepackage{algorithm}%http://rtlab.csie.ntu.edu.tw/resources/Latex/algorithms.pdf
\usepackage{algorithmic}
\theoremstyle{plain} \theorembodyfont{\upshape}

\newtheorem{definition}{\indent Definition}

\newtheorem{assumption}{\indent Assumption}

\newtheorem{problem}{\indent Problem}
\usepackage[usenames]{color}
\usepackage[svgnames]{xcolor}
\definecolor{DarkGreen}{rgb}{0,0.5,0}
\definecolor{DarkRed}{rgb}{0.75,0,0}
\usepackage{balance}
\usepackage[font=small]{caption}

\usepackage[normalem]{ulem}
\usepackage[capitalize]{cleveref}
\crefformat{equation}{(#2#1#3)}
\Crefformat{equation}{Equation~(#2#1#3)}
\Crefname{equation}{Equation}{Equations}

\IEEEoverridecommandlockouts                              % This command is only
\renewcommand{\hat}{\widehat}

\usepackage[footnote,printonlyused]{acronym}
\usepackage{scrextend}
\deffootnote[0.0em]{0em}{0em}
{\textsuperscript{\thefootnotemark}\,\enskip}

\title{\LARGE \bf
Active Sampling for Closed-loop Statistical Verification of \\Uncertain Nonlinear Systems}

\author{John F. Quindlen$^{1}$%
\thanks{$^{1}$Ph.D. Candidate, Department of Aeronautics and Astronautics, Massachusetts Institute of Technology (MIT).}%
\and Ufuk Topcu$^{2}$\thanks{$^{2}$Assistant Professor, Department of Aerospace Engineering and Engineering Mechanics, University of Texas at Austin}%
\and Girish Chowdhary$^{3}$\thanks{$^{3}$Assistant Professor, Departments of Agricultural \& Biological Engineering and Aerospace Engineering, University of Illinois Urbana-Champaign}%
\and Jonathan P.\ How$^{4}$\thanks{$^{4}$Richard C.\ Maclaurin Professor of Aeronautics and Astronautics, Laboratory for Information and Decision Systems (LIDS), MIT}%
}

\begin{document}
\raggedbottom 
\allowdisplaybreaks
 
\maketitle
\thispagestyle{empty}
\pagestyle{empty}

%%%%%%%%%%%%%%%%%%%%%%%%%%%%%%%%%%%%%%%%%%%%%%%%%%%%%%%%%%%%%%%%%%%%%%%%%%%%%%%%
\begin{abstract}
Increasingly demanding performance requirements for dynamical systems motivates the adoption of nonlinear and adaptive control techniques.  One challenge is the nonlinearity of the resulting closed-loop system complicates verification that the system does satisfy the requirements at all possible operating conditions.  
This paper presents a data-driven procedure for efficient simulation-based, statistical verification without the reliance upon exhaustive simulations.  In contrast to previous work, this approach introduces a method for online estimation of prediction accuracy without the use of external validation sets.  
This work also develops a novel active sampling algorithm that iteratively selects additional training points in order to maximize the accuracy of the predictions while still limited to a sample budget.  Three case studies demonstrate the utility of the new approach and the results show up to a 50\% improvement over state-of-the-art techniques.
\end{abstract}

%%%%%%%%%%%%%%
\section{Introduction}
\input{sec_intro}

%%%%%%%%%%%%%%
\section{Problem Formulation}\label{s:prob}
\input{sec_prob}

%%%%%%%%%%%%%%
\section{Statistical Data-Driven Verification}\label{s:modeling}
\input{sec_approach}

%%%%%%%%%%%%%%
\section{Closed-Loop Statistical Verification}\label{s:active}
\input{sec_sampling}

%%%%%%%%%%%%%%
\section{Examples}\label{s:results}
\input{sec_results}

%%%%%%%%%%%%%%
\section{Conclusion}\label{s:conclude}
This paper presented new data-driven verification methods for simulation-based certificates of complex nonlinear systems.  
%Given a suitable simulation model of the system, it is relatively straightforward to perform simulations at various uncertainties and use these to predict the system's ability to satisfy performance requirements at all possible conditions.  Unlike previous analytical verification techniques, which relied upon the existence of known analytical functions to bound the system's trajectories, simulation-based methods make no such assumptions and apply to a wider class of systems.  The ultimate goal is to construct a statistical certificate that correctly classifies whether the closed-loop response at particular uncertainties will satisfy the requirements.  
In particular, we introduced a GP-based verification framework to predict the satisfaction of requirements over the full space of possible operating conditions.  Additionally, this new approach exploits the availability of continuous measurements to quantify prediction confidence without external validation sets.  In many applications, the simulations themselves can be computationally expensive to obtain; therefore, it is advantageous to minimize the number of simulations required to obtain accurate predictions.  The second contribution of the paper is closed-loop verification using binary classification entropy for active selection of future training simulations.  Using this strategy, the examples in Section \ref{s:results} demonstrated up to a 50\% improvement in prediction error over existing approaches for the same number of samples.  
 
Ultimately, data-driven verification procedures are intended for use within higher-level problems such as robust, nonlinear planning or controller optimization.  In those those problems, verification is performed on each candidate control policy in order to estimate their robustness and the generation of candidate control policies is often an iterative process.  Thus, the process would typically cycle through a large number of candidate control policies and it is infeasible to test all possible uncertainties for every candidate policy.  Closed-loop, data-driven verification aligns with those problems by providing the controls engineer with the best approximation of the robustness while restricted to a budget on the number of allowable simulations during each verification step of the iterative process.  Upcoming work\cite{Quindlen17c_Arxiv} has further developed the closed-loop verification procedures and extended them to stochastic systems.

\section*{Acknowledgments} This work is supported by the Office of Naval Research and Air Force Office of Scientific Research under grants ONR MURI N00014-11-1-0688 and  AFOSR FA9550-15-1-0146 .

%\XX{fix the capitalization in the references} %---JQ: changed to IEEE standards
\balance
%%%%%%%%%%%%%%%%%%%%%%%%%%%%%%%%%%%%%%%%%%%%%%%%%%%%%%%%%%%%%%%%%%%%%%%%%%%%%%%%
\bibliographystyle{IEEEtran}
%\bibliography{BIB_Jack/JQ_15,BIB_Jack/bib_JQ13,BIB_all/ACL_Publications}
\bibliography{acc_2018,BIB_all/ACL_Publications,BIB_all/ACL_all,BIB_all/ACL_bef2000}

%%%%%%%%%%%%%%%
%\section{Appendix}
%\input{sec_appendix}

\end{document}

%% file: jfq_includes.tex
\usepackage{amsmath}
\usepackage[sort,compress]{cite}
\usepackage{graphicx}
\usepackage{epstopdf}
\usepackage{subfigure}
\usepackage{units}
\usepackage{rotating}
\usepackage{graphicx,paralist}
\usepackage{wrapfig}
\usepackage{nomencl}
\usepackage{amssymb}

\renewcommand{\vec}[1]{\boldsymbol{\mathbf{#1}}} %Easier to bold greek letters?
\newcommand{\degree}{\ensuremath{^\circ}}

\newcommand{\fig}[1] {Figure~\ref{#1}}

\newcommand{\vtheta}{\vec{\theta}}

%% file: sec_intro.tex
As the demand for higher performance and/or efficiency grows, advanced nonlinear control techniques will be relied upon to achieve the increasingly more complex requirements associated with such demands.  While methods like model reference adaptive control (MRAC)\cite{Lavretsky13} and reinforcement learning (RL)\cite{Sutton_98} have demonstrated large improvements in performance and efficiency, a key challenge with complex control techniques is certifying that the closed-loop system can actually meet the requirements at all possible uncertainties.  The nonlinear (and possibly adaptive) nature of the controller partly contributes to this difficulty since the controller nonlinearities can result in drastically different trajectories given only slightly different operating conditions. 
%This difficulty certifying the system is due, in part, to the nonlinear (and possibly adaptive) nature of the controller, which can result in drastically different trajectories given only slightly different operating conditions. 

Various verification techniques have been developed to address this problem.  If closed-form differential equations of the closed-loop system are available, then it may be possible to construct analytical certificates\cite{Moore12_CDC,Prajna05_PhD} to provably verify the closed-loop system satisfies the necessary requirements under certain modeling assumptions.  While analytical proofs are extremely useful, they are difficult to implement on many systems.  Simulation-guided analytical methods\cite{Kapinski14_HSCC,Topcu08_PhD,Reist15} relax this difficulty, but are still restricted by certain assumptions and may be overly conservative due to the reliance upon specific analytical functions for proof construction\cite{Quindlen16_ACC}.  

Statistical verification techniques\cite{Clarke11_ATVA,Zhang14_ATVA} construct statistical certificates directly from simulations of the closed-loop system.  While these certificates do not suffer from the same limitations as proof-based techniques and apply to a wider range of systems, they are less absolute.  If the observed simulation data fails to adequately cover the entire space of possible uncertainties, then the accuracy of the certificate's predictions in those unobserved regions will be limited.  
However, it is prohibitively expensive to simply saturate the set of all possible uncertainties with a large number of simulations, particularly when the system requires higher-fidelity verification models.

This work couples efficient data-driven statistical verification\cite{Kozarev16_HSCC} with active sampling\cite{Settles12} to carefully select simulations in order to minimize utilization of the model while maximizing prediction accuracy.  
%Active sampling\cite{Settles12} methods are closed-loop procedures that iteratively construct a model and then select informative training locations.  
%In particular, continuous robustness evaluations such as signal temporal logic\cite{Maler04_FORMATS} (STL) are used to measure satisfaction of the performance requirements during the simulation trajectories.  These measurements are then used to train Gaussian process\cite{Ras06} (GP) regression models that predict the satisfaction of the requirements at unseen perturbations.  
In particular, Section \ref{s:modeling} introduces a Gaussian process (GP) verification framework to train GP-based prediction models on a small set of trajectory robustness measurements and estimate the satisfaction of the requirements at unseen perturbations.  In order to quantify the model's confidence in these predictions, the framework also includes a simple method for online computation of prediction confidence using cumulative distributions.  Section \ref{s:active} exploits this validation metric to form a closed-loop verification procedure that  iteratively selects informative training locations to improve prediction accuracy.  
As existing active learning selection metrics\cite{Settles12,Kremer14_DMKD,Zhang16_AAAI} are not ideally suited to this binary verification problem, we develop a new tailor-made selection metric based on binary classification entropy.  %A batch active sampling procedure incorporates this new selection metric and iteratively selects informative training locations for improved predictions.  
Results in Section \ref{s:results} demonstrate the new statistical verification framework's improvement in prediction error when applied to three nonlinear systems.

%The paper is structured as follows.  Section \ref{s:prob} introduces the deterministic verification problem while Section \ref{s:modeling} describes the construction of GP prediction models.   Section \ref{s:active} presents the closed-loop active sampling procedure to intelligently select informative samples and retrain the model.  These procedures  attempt to form the best training set to produce the most accurate statistical certificate given a budget on the number of samples.  Section \ref{s:results} demonstrates these processes on three nonlinear systems: two model reference adaptive controllers and an aircraft autopilot. 

%% file: sec_prob.tex
Consider the deterministic nonlinear system
\begin{equation}\label{eq:system}
	\vec{\dot{x}}(t) = f(\vec{x}(t), \vec{u}(t), \vtheta)
\end{equation}
subject to parametric uncertainties $\vtheta\in\mathbb{R}^p$, where $\vec{x}(t) \in \mathbb{R}^n$ is the state vector and $\vec{u}(t) \in \mathbb{R}^m$ is the control input vector.  The open-loop plant in \cref{eq:system} is said to be deterministic; given the same $\{\vec{x},\vec{u},\vtheta\}$, the output will be the same every time.  Additionally, since the objective is to verify the closed-loop system, the control inputs $\vec{u}(t)$ are assumed to be generated by a known, deterministic control policy $\vec{u}(t) = g(\vec{x}(t))$.

The resulting closed-loop dynamics are a function of the parametric uncertainties $\vtheta$. These may include uncertainties about initial state $\vec{x}(0)$ or uncertain parameters such as mass/inertia properties of a physical system.  Regardless of their source, parameters $\vtheta$ are treated as uncertain initial conditions that affect the state dynamics.  Although the exact values of the parameters may be unknown at run-time, they are assumed to fall within a known, bounded set $\Theta$.
\begin{assumption}\label{assu:compactSet}
	The set of all possible perturbations $\vtheta \in \Theta$ is a known, compact, uncountable set $\Theta \in \mathbb{R}^p$.  
\end{assumption}
In practice, Assumption \ref{assu:compactSet} is not overly restrictive as most physical systems will have known feasible bounds on the operating conditions that can be used as $\Theta$.  %For instance, a commercial airliner will only fly within well-defined altitude and weight \& balance limits.

The trajectory of the closed-loop system is given by $\Phi(\vec{x}(t)|\vec{x}_0,\vtheta)$ and defines the evolution of state vector $\vec{x}(t)$ over the time interval $t\in[0,T_f]$.  %This trajectory is a deterministic function of nominal initial state vector $\vec{x}_0$ and perturbations $\vtheta$.  
Here, the nominal initial state vector $\vec{x}_0$ is assumed to be fixed and known.  If the true initial condition $\vec{x}(0)$ is uncertain $(\vec{x}(0) \neq \vec{x}_0)$, then $\vec{x}(0)$ can be modeled as the combination of known $\vec{x}_0$ and corresponding elements of the unknown perturbations, i.e. $\vec{x}(0) = \vec{x}_0 + \vtheta$.  %Effectively, only terms that vary across different instantiations of the system are incorporated into $\vtheta$.  
The resulting trajectory $\Phi(\vec{x}(t)|\vec{x}_0, \vtheta)$ will ultimately determine whether the performance requirements are satisfied at a particular $\vtheta$ condition.

%%%%%%%%%%%%%%%%%%%
\subsection{Problem Description}
%The closed-loop system trajectory is expected to satisfy certain pre-specified performance criteria.  These criteria may be supplied by a wide variety of sources or experts and can include relatively straightforward concepts like stability or boundedness as well as more complex spatial-temporal requirements such as those specified in civil or military aviation regulations\cite{FAR25,MIL1797}.  
The closed-loop system trajectory is expected to satisfy certain pre-specified performance criteria.  These criteria may be supplied by a wide variety of sources and can include relatively straightforward concepts like stability or boundedness as well as more complex spatial-temporal requirements such as those in Section \ref{s:mrac3D}.
This work assumes continuous (non-binary) measurements indicate the robustness of the trajectory to the requirements.
\begin{assumption}\label{assu:STL}
	Scalar variable $y \in \mathbb{R}$ measures the robustness of trajectory $\Phi(\vec{x}(t)|\vec{x}_0,\vtheta)$ .  The sign of $y$ indicates satisfaction of the requirements ($y > 0$ signifies the trajectory satisfied the requirements, while $y\leq0$ indicates failure).
\end{assumption}
As fixed $\vec{x}_0$ and condition $\vtheta$ completely define the evolution of each trajectory, we write the measurements as $y(\vtheta)$ to emphasize robustness is an explicit function of $\vtheta$.

%These continuous measurements could be provided by any relevant source, but commonly would be produced by signal temporal logic (STL)\cite{Maler04_FORMATS}.  
Signal temporal logic (STL)\cite{Maler04_FORMATS} commonly, but not exclusively, provides these continuous measurements of robustness to the requirements. 
Signal temporal logic is a mathematical language for expressing the requirements as functions of logical predicates and boolean and/or temporal operators.  In comparison to other temporal logic frameworks, STL uniquely provides a continuous-valued robustness degree $\rho^{\varphi} \in \mathbb{R}$ to quantify the robustness of a trajectory with respect to requirement $\varphi$.  The availability of this $\rho^{\varphi}$ or a similar measurement to quantify a trajectory's robustness is central to the approach in this paper.  Similar work without the availability of non-binary measurements has been presented in \cite{Kozarev16_HSCC,Quindlen18_GNC}.  While it applies to a slightly larger class of systems, the approach suffers from many limitations that will be discussed later in this paper.

Even though continuous $y(\vtheta)$ measures the trajectory's robustness, the satisfaction of the requirements for deterministic systems is purely binary: at given condition $\vtheta$, the corresponding trajectory will either satisfy the requirements or it will not.  Due to the binary aspect of the problem, the feasible set $\Theta$ can be segmented into two unique sets.
\begin{definition}\label{def:ros}
	The \emph{region of satisfaction} $\Theta_{sat}$ contains all $\vtheta \in \Theta$ for which the resulting trajectory satisfies the performance requirements,
	\begin{equation}
		\Theta_{sat} = \Big\{ \vtheta \in \Theta: y(\vtheta) > 0\Big\}.
	\end{equation}
\end{definition}
\begin{definition}\label{def:rof}
	The \emph{region of failure} $\Theta_{fail}$ contains all $\vtheta \in \Theta$ for which the resulting trajectory fails to satisfy the requirements, 
	\begin{equation}
		\Theta_{fail} = \Big\{ \vtheta \in \Theta: y(\vtheta) \leq 0\Big\}.
	\end{equation}
\end{definition}
It is assumed $\Theta_{sat},\Theta_{fail} \neq \emptyset$.  By construction, $\Theta_{sat} \cup \Theta_{fail} = \Theta$ and $\Theta_{sat} \cap \Theta_{fail} = \emptyset$.  While the conditions for membership in sets $\Theta_{sat}, \Theta_{fail}$ are known, the sets themselves are unknown in advance; it is not clear whether an arbitrary $\vtheta$ belongs to $\Theta_{sat}$ or $\Theta_{fail}$ without running a simulation.  The verification goal is to compute an estimated region of satisfaction, $\hat{\Theta}_{sat}$. 
\begin{problem}\label{prob:cert}
	Given a deterministic closed-loop system and measurements of requirement satisfaction, compute an estimated region of satisfaction $\hat{\Theta}_{sat}$, with $\hat{\Theta}_{fail} = \Theta \setminus \hat{\Theta}_{sat}$.
\end{problem}

%% file: sec_approach.tex
Problem \ref{prob:cert} can be viewed as a \emph{binary classification} problem: predict whether $\vtheta$ belongs to $\Theta_{sat}$ or $\Theta_{fail}$.  Previous work\cite{Kozarev16_HSCC,Quindlen18_GNC} constructed $\hat{\Theta}_{sat}$ and $\hat{\Theta}_{fail}$ through support vector machines (SVM) classification models.  While the results demonstrated the ability of the SVM-based approach to product accurate estimates, there was no way to efficiently quantify confidence in the predictions online without relying upon external validation datasets.  This mainly evolves out of the fact binary classification models only utilize binary evaluations rather than continuous $y(\vtheta)$.  SVM extensions such as Platt scaling\cite{Platt99_ALMC} produce approximate measures of confidence without validation sets, but these confidence estimates rely upon artificial approximations of $y(\vtheta)$ output from the SVM model rather than the true $y(\vtheta)$ values themselves.

Although Assumption \ref{assu:STL} indicates the sign of $y(\vtheta)$ can always convert continuous $y(\vtheta)$ into binary evaluations, it discards valuable information.  In comparison to binary measurements, $y(\vtheta)$ also quantifies ``just how close'' a satisfactory trajectory came to failure or ``how bad'' an unsatisfactory one performed.  The main contribution of this section is a new verification framework that trains a regression model in order to avoid losing the information stored in the robustness degree.  Even though regression models replace SVMs, the problem is still fundamentally binary classification.  To emphasize this, we label the new approach \emph{regression-based binary verification}.  With the extra information, regression-based binary verification also simultaneously provides explicit measures of prediction confidence online without any additional validation dataset.

%%%%%%%%%%%%%%%%%%%%%%%
\subsection{Gaussian Process Regression Model}
The noise-free GP regression model is constructed from a training dataset $\mathcal{L}$ of initial observations.  This training dataset consists of $N$ pairs of $\vtheta$ values and their measurements $y(\vtheta)$.  The set of $N$ observed perturbation conditions is $\mathcal{D} = \{\vtheta_1, \vtheta_2, \hdots \vtheta_N\}$ while the measurements are grouped in vector $\vec{y} = [y(\vtheta_1), y(\vtheta_2),\hdots y(\vtheta_N)]^T$.  The training dataset is then $\mathcal{L} = \{\mathcal{D},\vec{y}\}$.
At its core, a GP defines a distribution over possible functions that predicts the value of $y(\vtheta)$ in unobserved regions of $\Theta$.  A more in-depth discussion of GPs is found in \cite{Rasmussen06}.  Assuming a zero-mean prior, the end result of the training process is a Gaussian posterior predictive distribution at arbitrary location $\vtheta_*$ with mean $\mu(\vtheta_*)$ and covariance $\Sigma(\vtheta_*)$,
\begin{align}\label{eq:GP}
	\mu(\vtheta_*) & = \vec{K}_*^T \vec{K}^{-1}\vec{y} \notag \\
	\Sigma(\vtheta_*) & = \kappa(\vtheta_*,\vtheta_*) - \vec{K}_*^T \vec{K}^{-1} \vec{K}_*.
\end{align}
Term $\vec{K}_*$ is a $N \times 1$ vector of $\kappa(\vtheta_*,\vtheta_i) \ \forall i=1:N$ while $\vec{K}$ is a $N \times N$ matrix of $\kappa(\vtheta_i,\vtheta_j) \ \forall i,j=1:N$.  
Although there are many possible options for kernel function $\kappa$, this paper uses the squared exponential kernel with automatic relevance determination (ARD) and kernel hyperparameters $\psi$.  In comparison to the isotropic squared exponential kernel, the ARD kernel enables the hyperparameters to vary and thus deemphasize elements of $\vtheta$ with minimal impact upon $y(\vtheta)$ or emphasize those with high sensitivity.  

%%%%%%%%%%%%%%%%%%%%%%%%%%%
The model presented in \cref{eq:GP} is the standard GP model.  Although this is by far the most widely-used formulation, it has been shown to have difficulty with higher-dimensional systems.  More complicated extensions\cite{Kandasamy15_ICML,Hoang15_ICML} decompose the GP into the sum of additive models or find sparse approximations of the full GP.  For simplicity, we focus on verification using standard GPs, but these more complex extensions could be readily incorporated into our approach.

%%%%%%%%%%%%%%%%%%%%%%%
\subsection{Importance of Hyperparameter Optimization}\label{ssec:hypOpt}
The choice in hyperparameters $\psi$ greatly affects the GP regression model and ultimately the posterior predictive distributions that define $\hat{\Theta}_{sat},\hat{\Theta}_{fail}$.  Interestingly enough, the majority of works related to GP-based verification\cite{Chen16_CDC,Zhang16_AAAI,Gotovos13_IJCAI,Desautels14_JMLR} make little-to-no mention of the choice of hyperparameters or assume the optimal choice of hyperparameters that perfectly replicates the underlying function for $y(\vtheta)$ is known and fixed.  For Problem \ref{prob:cert}, this assumption cannot be made as little is known about the distribution of the robustness degree over $\vtheta$ until after simulations are performed.  Instead, the hyperparameters must be chosen with only the current available information, training dataset $\mathcal{L}$.
 
The likelihood distribution of the hyperparameters with respect to the training set $\mathcal{L}$ is given by $\mathbb{P}(\psi|\mathcal{L})$.  The main issue is that there is no closed-form solution to compute $\mathbb{P}(\psi|\mathcal{L})$ and sampling-based approximations or Markov Chain Monte Carlo (MCMC) methods are intractable.  
Instead, maximum likelihood estimation (MLE) approximates the hyperparameters with local optimums.  These methods efficiently compute a local maximum using steepest descent methods or gradient-based methods \cite{Rasmussen06}.

%%%%%%%%%%%%%%%%%%%%%%%
\subsection{Binary Predictions with Prediction Confidence}\label{s:confidence}
The posterior mean $\mu(\vtheta)$ provides the basis for the binary predictions.  The mean represents the expected value of the robustness degree, $\mu(\vtheta) = \hat{y}(\vtheta)$, and defines $\hat{\Theta}_{sat}$,$\hat{\Theta}_{fail}$ using the equivalent of Definitions \ref{def:ros} and \ref{def:rof} with $\hat{y}(\vtheta)$.  Unlike earlier work with SVM-based predictions\cite{Kozarev16_HSCC,Quindlen18_GNC}, the GP-based model also quantifies the prediction confidence without requiring an external validation dataset.  The Gaussian cumulative distribution function (CDF) associated with the GP output determines the probability of satisfaction at each $\vtheta$ location,
\begin{equation}\label{eq:probSat}
	\mathbb{P}(y(\vtheta)>0|\mathcal{L},\psi) = \mathbb{P}_{+}(\vtheta) = \frac{1}{2} +\frac{1}{2} \text{erf}\Big(\frac{\mu(\vtheta)}{\sqrt{2 \Sigma(\vtheta)}}\Big),
\end{equation}
and indicates the likelihood of misclassification error there.

The choice of hyperparameters will also affect the CDF describing the probability of requirement satisfaction.  If the true hyperparameters, labeled $\psi^*$, are known, then the actual probability $\mathbb{P}(y(\vtheta)>0|\mathcal{L}) = \mathbb{P}(y(\vtheta)>0|\mathcal{L},\psi^*)$.  However, in the more likely case when $\psi^*$ is not known, the probability of satisfaction is the marginal likelihood over all possible $\psi$,
\begin{equation}\label{eq:probSat2}
	\mathbb{P}(y(\vtheta)>0|\mathcal{L}) = \int \mathbb{P}(y(\vtheta)>0|\mathcal{L},\psi) \mathbb{P}(\psi|\mathcal{L})  d\psi \ .
\end{equation}
Since this integral requires $\mathbb{P}(\psi|\mathcal{L})$, the computation of the total probability \cref{eq:probSat2} is intractable as well and the MLE approximation can be used in its place.

%% file: sec_sampling.tex
Data-driven verification must contend with the dueling objectives of saturating $\Theta$ for better accuracy and minimizing $|\mathcal{L}|$ due to computational costs.  This work assumes the primary source of computational cost is the simulation model.  In many applications, the model fidelity required for accuracy verification is quite high.  For instance, aircraft simulations typically require a full nonlinear flight simulation model to capture the interactions of flight dynamics, actuator saturation, and control and their effects on performance.
Even with lower-fidelity models, verification may be part of a larger process such as robust nonlinear control design, which restricts the number of simulations that can be feasibly allocated to each candidate controller.
To model these practical limits, we assume the computational budget restricts the number of simulations  allocated to the verification procedure.  
\begin{assumption}\label{assu:budget}
	The computational budget manifests as a cap on the number of simulations, $N_{total}$.
\end{assumption}
As the space $\Theta$ is uncountable, the approach assumes there exists an extremely fine discretization $\Theta_d$ which approximates $\Theta$.  Even with this finite discretization, $|\Theta_d| \gg N_{total}$ and it is impossible to saturate $\Theta_d$ with training locations $\mathcal{D}$.  The remaining, unobserved sample locations are labeled $\mathcal{U} = \Theta_d \setminus \mathcal{D}$.

\subsubsection{Active Sampling}
With $\mathcal{L}$ limited to $N_{total}$ samples, it is crucial to select the most informative training set.  As expected, the difficulty lies in the fact that the ideal training set is unknown apriori.  Rather than \textit{passively} select training datapoints, either randomly or in a structured grid, active learning selects informative training samples in an iterative manner.
Active learning\cite{Settles12} describes a wide variety of different procedures, all of which attempt to identify the ``best'' samples to obtain in order to improve a statistical model.  Depending on procedure's objective, the definition of the best sample will change, even for the same exact model.  
One of the most general approaches is the expected model change (EMC) metric\cite{Kremer14_DMKD} used by our prior work in closed-loop verification with binary measurements\cite{Quindlen18_GNC}.  Though mostly applied to SVM models, EMC can be extended to GP regression models.  The  ``best'' sample $\overline{\vtheta}$ is the point most likely to induce the largest expect change in the model, meaning $\overline{\vtheta} = \text{argmin }|\mu(\vtheta)|$.  Conversely, the most common GP-based approach is variance reduction\cite{Zhang16_AAAI}.  While various approximations or derivations exist\cite{Gotovos13_IJCAI,Desautels14_JMLR}, the end goal is to reduces the posterior variance of the GP model.  Since it is comparatively expensive to calculate the change in posterior variance, the most common approximation is to select $\overline{\vtheta} = \text{argmax }\Sigma(\vtheta)$\cite{Zhang16_AAAI}.  
Regardless of their efficiency or accuracy, none of these approaches directly address the fundamental verification objective: to predict whether $y(\vtheta) > 0$ or not.  

Closed-loop statistical verification introduces a new selection metric to identify informative sample locations: binary classification entropy.  Unlike variance methods, which may also use the term ``entropy''\cite{Zhang16_AAAI}, binary classification entropy exploits the probability $\mathbb{P}(y(\vtheta)>0)$,
\begin{equation}\label{eq:binaryEntropy}
\begin{aligned}
	H(\vtheta|\mathcal{L},\psi) & = -\bigg(\mathbb{P}_{+}(\vtheta) \ \text{log}_2 \mathbb{P}_{+}(\vtheta) \\
	& \ \ \ \ \ + (1 -\mathbb{P}_{+}(\vtheta)) \ \text{log}_2 (1-\mathbb{P}_{+}(\vtheta))\bigg) ,
\end{aligned}
\end{equation}
rather than the entropy of $\Sigma(\vtheta)$.

Ideally, the best possible sample location $\overline{\vtheta}$ would be to minimize the total posterior entropy.  Unfortunately, it is impractical to compute the expected posterior change in entropy for the same reason as variance reduction methods\cite{Zhang16_AAAI}.  Instead, closed-loop statistical verification selects the sample location with the largest reduction in \textit{local} entropy
\begin{align}\label{eq:feasibleSeq}
	\overline{\vtheta} & =  \underset{\vtheta'}{\text{argmax }}\Bigg( H(\vtheta'|\mathcal{L},\psi) -  H(\vtheta'|\mathcal{L}^+,\psi)\Bigg) \notag \\
		& = \underset{\vtheta'}{\text{argmax }} H(\vtheta'|\mathcal{L},\psi) .
\end{align}
Once a sample is taken, the entropy at that location is 0; therefore, the point with the largest magnitude of classification entropy would have the largest reduction.  Note that the binary classification entropy is strictly non-negative.

%%%%%%%%%%%%%%%%
\subsubsection{Batch Sampling}\label{ssec:batchSampling}
While a sequential sampling procedure based upon \cref{eq:feasibleSeq} will correctly guide the selection of sample points as intended, it requires a large amount of computational effort to continuously recompute the GP after each iteration.  
Batching sampling presents one of the most practical methods for reducing the retraining cost - select multiple samples between retraining steps.  Batch active sampling methods\cite{Settles12,Kremer14_DMKD,Gotovos13_IJCAI,Desautels14_JMLR} lower the training cost and exploit any inherent parallel computing capabilities.  These approaches select $M$ datapoints at once and perform their corresponding simulations in parallel.  The challenge is to ensure adequate diversity in the chosen batch set $\mathcal{S}$ in order to avoid the selection of redundant trajectories.

The second major contribution of this paper is a batch active sampling extension of \cref{eq:feasibleSeq} for closed-loop statistical verification.  
This approach utilizes importance weighting, essentially modified importance sampling\cite{Rubinstein08}, to efficiently select samples using only the current entropy.
Unlike variance-based procedures, this approach freezes the information and selects the entire batch of $M$ samples without any intermediate retraining.  The current entropy $H(\vtheta|\mathcal{L},\psi)$ is used to construct a probability distribution from which samples of $\vtheta$ can be chosen, where regions with a large magnitude of entropy will have a higher probability of selection,
\begin{equation}
	\mathbb{P}_H(\vtheta) = \frac{1}{Z_H}H(\vtheta|\mathcal{L},\psi) \ \ \  \text{where } Z_H = \sum_{i=1}^{|\vtheta_d|}H(\vtheta_i|\mathcal{L},\psi)  .
\end{equation}
Although pure random sampling based upon $\mathbb{P}_H(\vtheta)$ generates samples that are clustered in regions with large probability, it does not ensure adequate diversity in batch $\mathcal{S}$.  Particularly when batch size $M$ is low, it is desirable to spread samples out across regions with similarly-high probability.  The problem is that the randomly-selected samples may inadvertently cluster into one area rather than spread out, thus lacking diversity.  In order to address this issue, importance-weighting can be augmented with random matrix theory methods to encourage diversity in the samples while still steering samples towards regions with high probability/entropy.  

The central tool to encourage diversity is Determinantal Point Processes (DPPs)\cite{Kulesza12_FTML}.  DPPs are probabilistic models for efficient sampling that penalize correlations in the samples and therefore can be used to discourage similarities between samples.  Algorithm \ref{alg:entropyRMT} presents the batch active sampling procedure using DPPs.  Each iteration's DPP forms a $\vtheta$-correlation matrix $L$ from a large number $(M_T \approx 1000)$ of $\vtheta$ locations randomly generated according to $\mathbb{P}_H(\vtheta)$.  Since $M_T$ must be so large to approximate $\mathbb{P}_H(\vtheta)$, but the batch size $M$ is generally small, Algorithm \ref{alg:entropyRMT} implements a variant of DPPs called k-DPPs\cite{Kulesza11_ICML} to correctly obtain $M \ll M_T$ samples in a sequential manner.  Once a sample has been selected, the remaining samples are weighted towards points with corresponding eigenvectors of $L$ that are orthogonal to that recently-selected sample.  The finished batch $\mathcal{S}$ disperses the selected $\vtheta$ locations among regions of high entropy/probability with less redundancy.  It is important to note that while the k-DPP requires $M_T$ random samples from $\mathbb{P}_H(\vtheta)$, it does not actually perform simulations at all those locations and only introduces a small additional cost over baseline importance sampling.

\begin{algorithm}[t]
%\vspace{0.2in}
\caption{\strut Batch active sampling using k-DPP.}
\label{alg:entropyRMT}
\begin{algorithmic}[1]
	\STATE \textbf{Input:} training set $\mathcal{L}$, available sample locations $\mathcal{U}$, $T$ batches, batch size $M$
	\STATE \textbf{Initialize:} GP regression model, entropy $H(\vtheta|\mathcal{L},\psi)$
	\FOR{$i=1:T$}
		\STATE{\textbf{Initialize:} empty set $\mathcal{S} = \emptyset$}
		\STATE{Transform $H(\vtheta|\mathcal{L},\psi)$ into $P_H(\vtheta)$}
		\STATE{Randomly generate $M_T$ locations, construct k-DPP}
		\STATE{Obtain $M$ samples according to k-DPP, add to $\mathcal{S}$}
		\STATE{Run simulations $\forall \vtheta \in \mathcal{S}$, obtain measurements $\vec{y}_{\mathcal{S}}$}
		\STATE{Add $\{ \mathcal{S},\vec{y}_{\mathcal{S}}\}$ to $\mathcal{L}$, remove $\mathcal{S}$ from $\mathcal{U}$}
		\STATE{Retrain regression model $\mathcal{GP}$ with new $\mathcal{L}$}
	\ENDFOR
	\STATE \textbf{Return:} predicted $\hat{\Theta}_{sat}, \hat{\Theta}_{fail}$
\end{algorithmic}
\end{algorithm}

%% file: sec_results.tex
This section demonstrates Algorithm \ref{alg:entropyRMT} on three example problems of interest.  Two of the examples consider adaptive control systems.  The third examines a lateral-directional autopilot and demonstrates closed-loop verification on a very complex system with nonlinear open-loop dynamics, control saturation, sensor models, and switching logic.  %Each example also highlights one aspect of closed-loop verification for further discussion in addition to total misclassification (prediction) error.

%%%%%%%%%%%%%%%
\subsection{Example 1: Model Reference Adaptive Control}\label{s:mrac}
The first example is a concurrent learning model reference adaptive control (CL-MRAC) system\cite{Chowdhary10_CDC}.  This problem has been examined before with other approaches, namely proof-based techniques that produced overly-conservative analytical certificates\cite{Quindlen16_ACC} and data-driven certificates with binary measurements\cite{Kozarev16_HSCC,Quindlen18_GNC}.  The following subsection will demonstrate that entropy-based active sampling procedures can produce accurate data-driven certificates of satisfaction without relying upon large training datasets.  

In this problem, the open-loop plant dynamics are subject to two uncertain parameters, $\vtheta = [\theta_1,\theta_2]^T$.  The goal of the system is to track a linear reference model with a desirable second-order response (state $\vec{x}_m(t)$) over a 40 second trajectory.  The tracking error, the distance between the actual and reference trajectories, is given by $\vec{e}(t) = \vec{x}_m(t) - \vec{x}(t)$.  The adaptive controller estimates the uncertain parameters online and augments the nominal control input with an additional component. %$\vec{u}_{ad}(t) = \vec{\hat{\theta}} \cdot \vec{x}(t)$.
As a result of the adaptive law and the uncertainties, the closed-loop dynamics are nonlinear and difficult to analyze.  

While the CL-MRAC adaptive law ensures the asymptotic convergence of the trajectory, it is difficult to prove boundedness of the transient errors.  In particular, the performance requirement considered in this example is that the tracking error component $e_1(t)$ remains bounded within the unit ball at all times, %written in STL format as
\begin{equation}\label{eq:STLmrac}
	\varphi_{bound} = \Box_{[0, 40]} \ (1 - |e_1[t]| \geq 0) .
\end{equation}
The performance measurement $y(\vtheta)$ is the STL robustness degree $\rho^{\varphi}(\vtheta)$ given by
\begin{equation}
	\rho^{\varphi}(\vtheta) = \underset{t' \in [0,40]}{\text{min }} \rho^{\varphi_{bound}}[t'](\vtheta).
\end{equation}
The goal of the active sampling procedures is to measure $y(\vtheta) = \rho^{\varphi}(\vtheta)$ at various training locations and predict $\mathbb{P}(y(\vtheta)>0)$ at unobserved $\vtheta \in \Theta_d$.  For this example, $\Theta_d$ covers the space between $\theta_1: [-10,10]$ and $\theta_2: [-10,10]$ with a total of 40,401 possible sample locations.  It is unlikely $N_{total} = 40,401$ in most computationally-constrained scenarios, so the samples must be carefully selected to improve the predictions.

\begin{figure}[!]
	\centering
	{\includegraphics[width=.79\columnwidth]{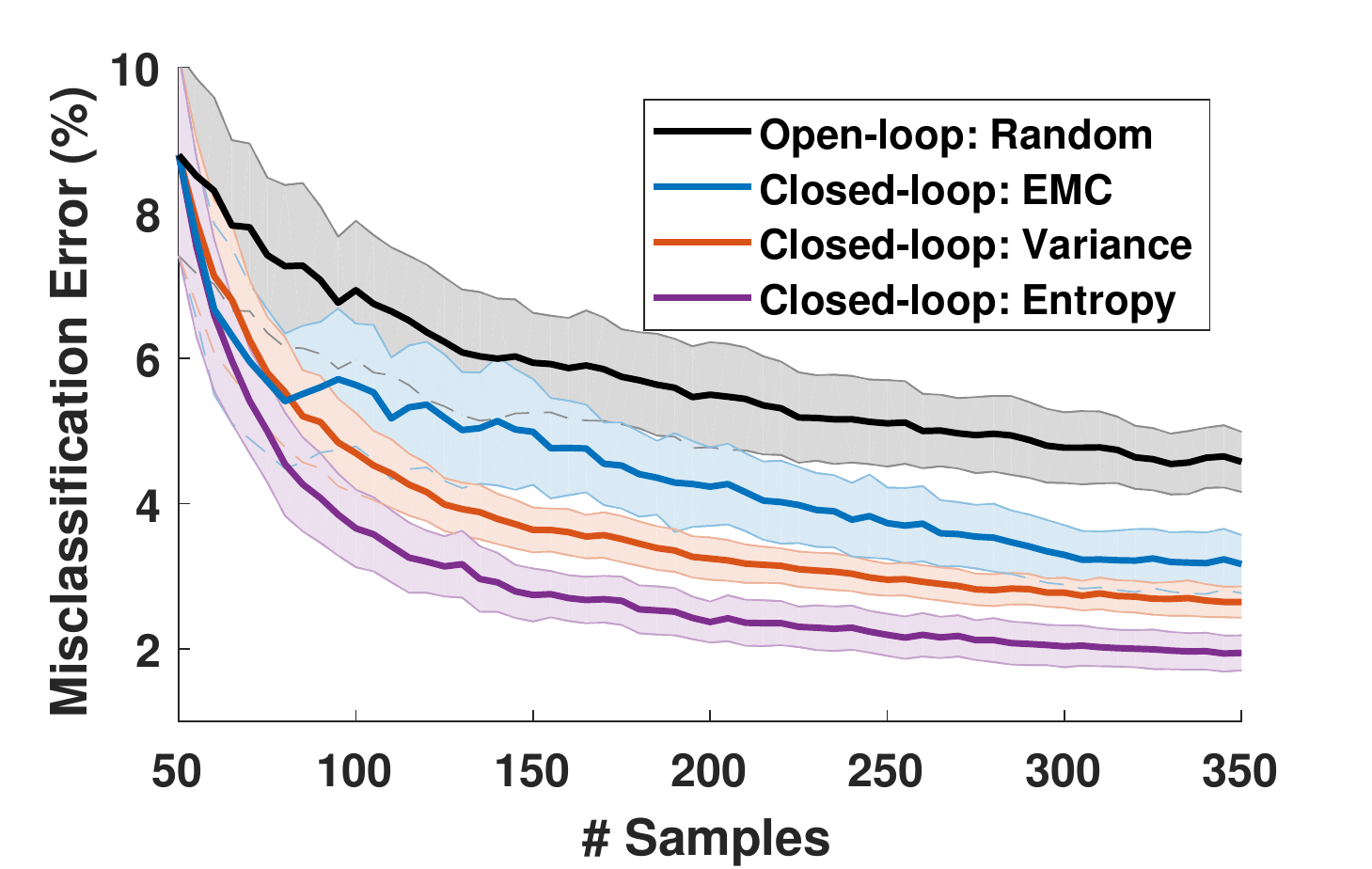}}
	\caption{(Example 1) Total misclassification error at batch size $M=5$ over 100 randomly-initialized runs.  Standard deviation intervals correspond to 0.5$\sigma$ bounds for easier viewing.}
 	\label{f:mrac1a}
	\vspace{-0.1in}
\vspace{0.2in}
	\centering
	{\includegraphics[width=.79\columnwidth]{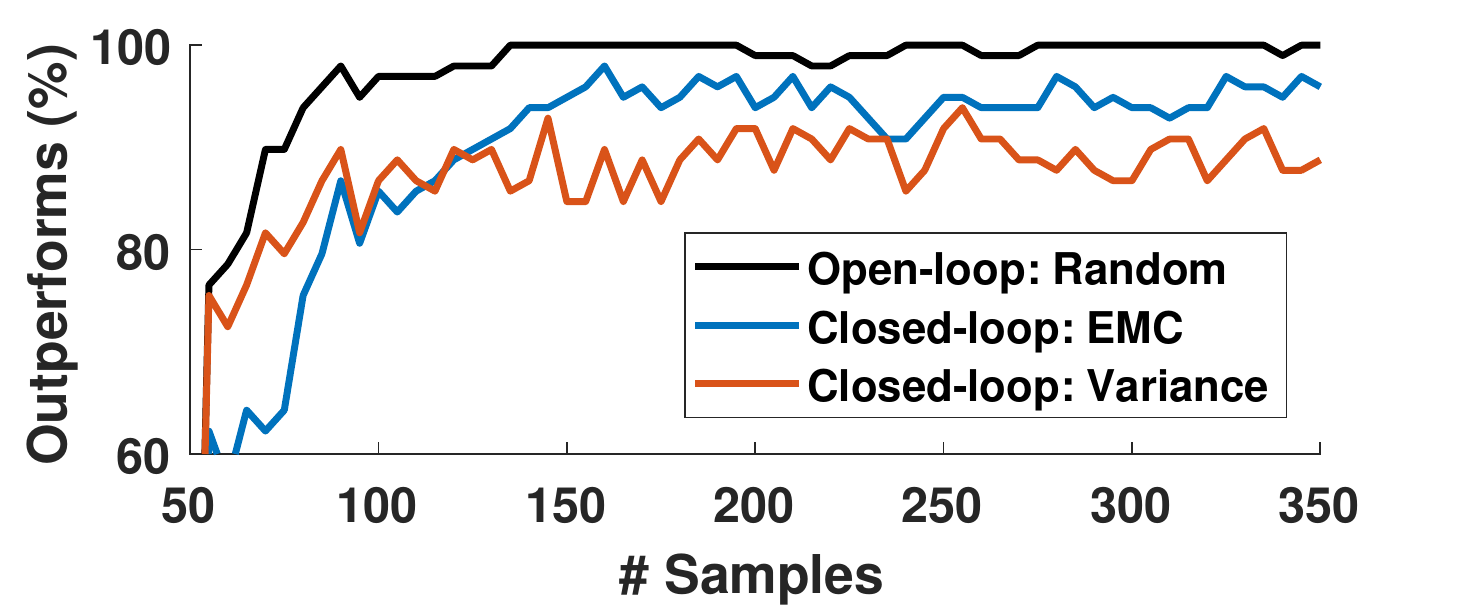}}
	\caption{(Example 1) Percentage of the 100 runs where the entropy-based procedure directly matches or outperforms the competing approach given the same initial training dataset and GP model.}
 	\label{f:mrac1b}
	\vspace{-0.1in}
\end{figure}

Figure \ref{f:mrac1a} compares the verification performance of Algorithm \ref{alg:entropyRMT} against batch versions of the existing EMC and variance-based methods as well as passive, random sampling.  The procedures all start with the same 100 initial training sets of 50 randomly-selected samples and operate in batches of size $M=5$ for a total of $N_{total} = 350$ points.  Entropy-based sampling outperforms all the competing techniques in average misclassification error.  At the conclusion of the process after $|\mathcal{L}|=350$ training points, Algorithm \ref{alg:entropyRMT} demonstrates a 24.33\% average improvement in misclassification error over the closest competitor, variance-based sampling, as well as a 34.6\% improvement over the EMC selection metric used by closed-loop verification with binary evaluations\cite{Quindlen18_GNC}.  Since the four strategies all start from the same initial training set and GP model in each of the 100 runs, the results also directly compare their performance in each of those test cases.  \fig{f:mrac1b} plots the ratio of the 100 runs where the misclassification rate of Algorithm \ref{alg:entropyRMT} either matches or exceeds that of the indicated competing strategy.  The new active sampling method matches or outperforms the top-competing variance-based method in 88\% of the runs.

\begin{figure}[!]
	\centering
	{\includegraphics[width=.79\columnwidth]{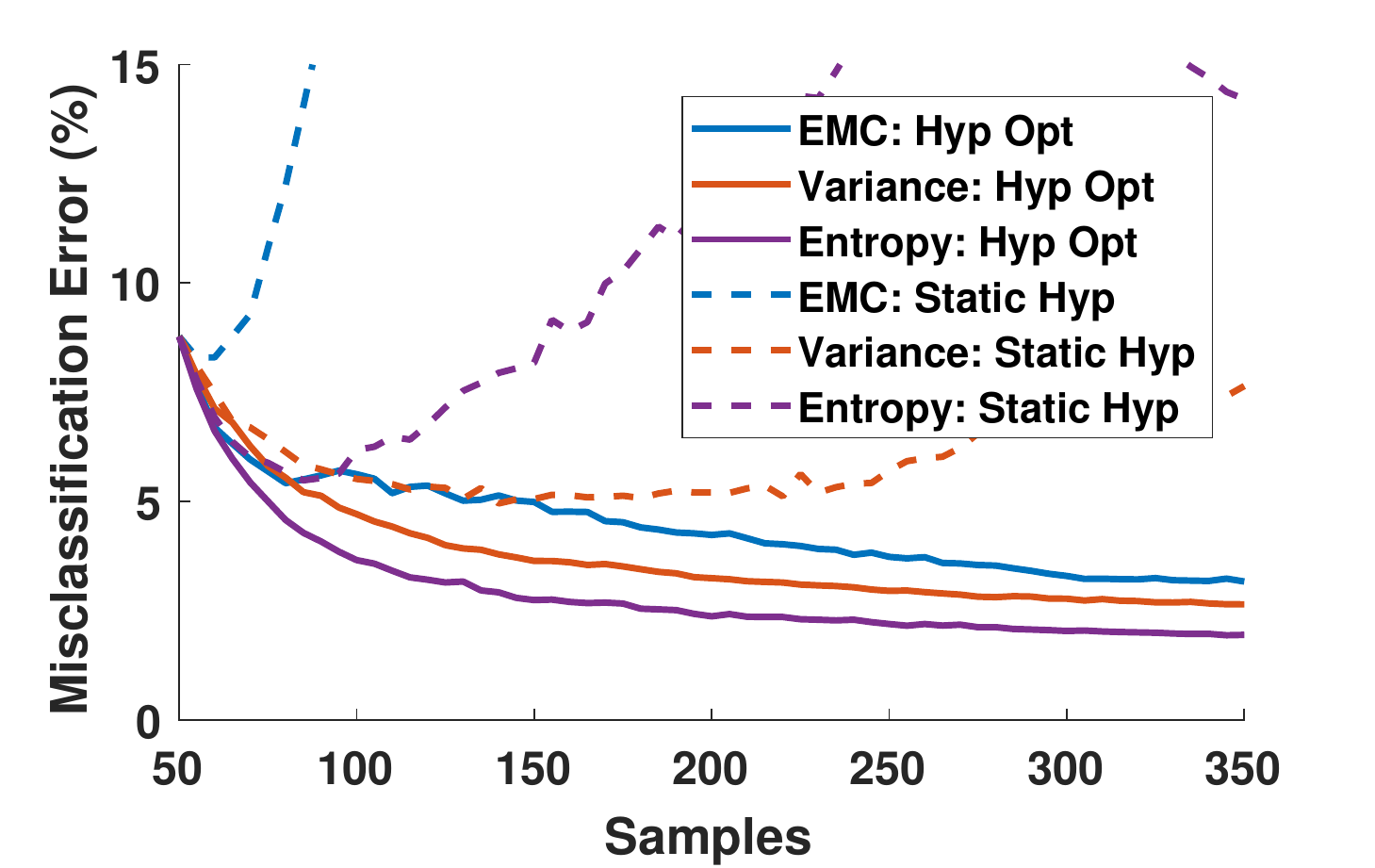}}
	\caption{(Example 1) Average misclassification error of closed-loop verification procedures with hyperparameter optimization versus the same algorithms with static (constant) hyperparameters.}
 		\label{f:mrac1c}
	\vspace{-0.1in}
\end{figure}

\fig{f:mrac1c} illustrates the importance of hyperparameter optimization when nothing is known about the correct choice of hyperparameters.  If the active sampling algorithms naively fix the hyperparameters to their initial values and never update them, then the prediction errors may actually \emph{increase} with additional simulation data.  This mainly serves to highlight the importance of the choice of hyperparameters and potential issues with sampling approaches\cite{Gotovos13_IJCAI,Desautels14_JMLR} that assume fixed hyperparameters.

%\begin{figure}[!]
%	\centering
%	{\includegraphics[width=.79\columnwidth]{figures/mrac2D-final/error95_batch5.pdf}}
%	\caption{(Example 1) Misclassification error within the 95\% prediction confidence levels. Unlike the total misclassification error, this only counts prediction errors for points with high confidence. Standard deviation intervals correspond to 0.5$\sigma$ bounds for easier viewing.}
% 		\label{f:mrac1d}
%	\vspace{-0.1in}
%\end{figure}

%%%%%%%%%%%%%%%%
\subsection{Example 2: Adaptive Control with Complex Temporal Requirements}\label{s:mrac3D}
The second example verifies the same CL-MRAC system from Section \ref{s:mrac}, but with a different set of performance requirements.  In this example, the adaptive system must satisfy three different signal temporal logic specifications, i.e. $\varphi = \varphi_1 \wedge \varphi_2 \wedge \varphi_3$.  These three specifications are
\begin{equation}
\begin{aligned}
	&\ \ \varphi_1 = \Diamond_{[2,3]} \ (x_1[t] - 0.7 \geq 0) \ \ \wedge \\
	& \ \ \ \ \ \ \ \ \ \ \ \ \ \ \ \ \Diamond_{[2,3]} \ (1.3 - x_1[t] \geq 0), \\
	&\ \ \varphi_2 = \Diamond_{[12,13]} \ (x_1[t] - 1.1 \geq 0) \ \ \wedge \\
	& \ \ \ \ \ \ \ \ \ \ \ \ \ \ \ \ \Diamond_{[2,3]} \ (1.7 - x_1[t] \geq 0), \\
	\text{and }& \ \ \varphi_3 = \Box_{[22.4,22.6]} \ (x_1[t] + 1.6 \geq 0) \ \ \wedge \\
	& \ \ \ \ \ \ \ \ \ \ \ \ \ \ \ \ \Box_{[22.4,22.6]} \ (-1.2 - x_1[t] \geq 0)\ .
\end{aligned}
\end{equation}
The total robustness measurement for each trajectory is then
\begin{equation}
	 y(\vtheta) = \text{ min }\{\rho^{\varphi_1}(\vtheta), \ \rho^{\varphi_2}(\vtheta), \ \rho^{\varphi_3}(\vtheta)\}.
\end{equation}

This example considers the same two uncertain parameters $(\theta_1,\theta_2)$ from Section \ref{s:mrac}, but adds a third parameter $\theta_3$ for uncertain initial state $x_1(0)$.  The new sampling set $\Theta_d$ covers the space of $\theta_1: [-5,5]$, $\theta_2:[-5,5]$, and $\theta_3:[-1,1]$ with a grid of 214,221 possible sample locations.  Just as with the previous example, the four verification procedures start with the same initial training dataset for each of the 100 runs.  In this example, these initial sets consist of 100 random $\vtheta$ locations and the algorithms operate in batches of $M=10$ points until they reach a total of $N_{total} = 1,050$ samples.

\begin{figure}[!]
	\centering
	{\includegraphics[width=.79\columnwidth]{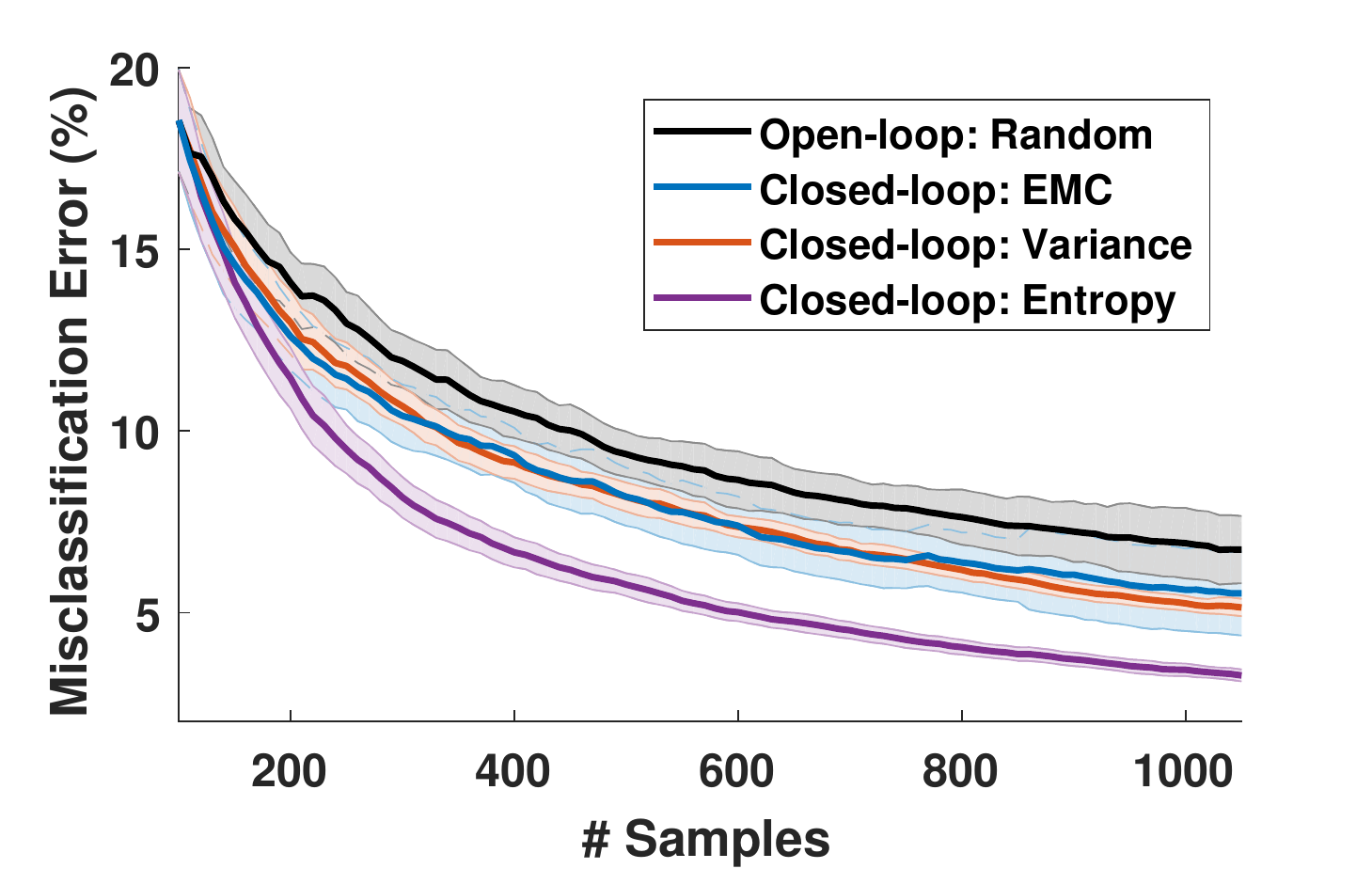}}
	\caption{(Example 2) Total misclassification error at batch size $M=10$ over 100 randomly-initialized runs.  Standard deviation intervals correspond to 0.5$\sigma$ bounds.}
 	\label{f:mrac2a}
	\vspace{-0.1in}
\vspace{0.2in}
	\centering
	{\includegraphics[width=.79\columnwidth]{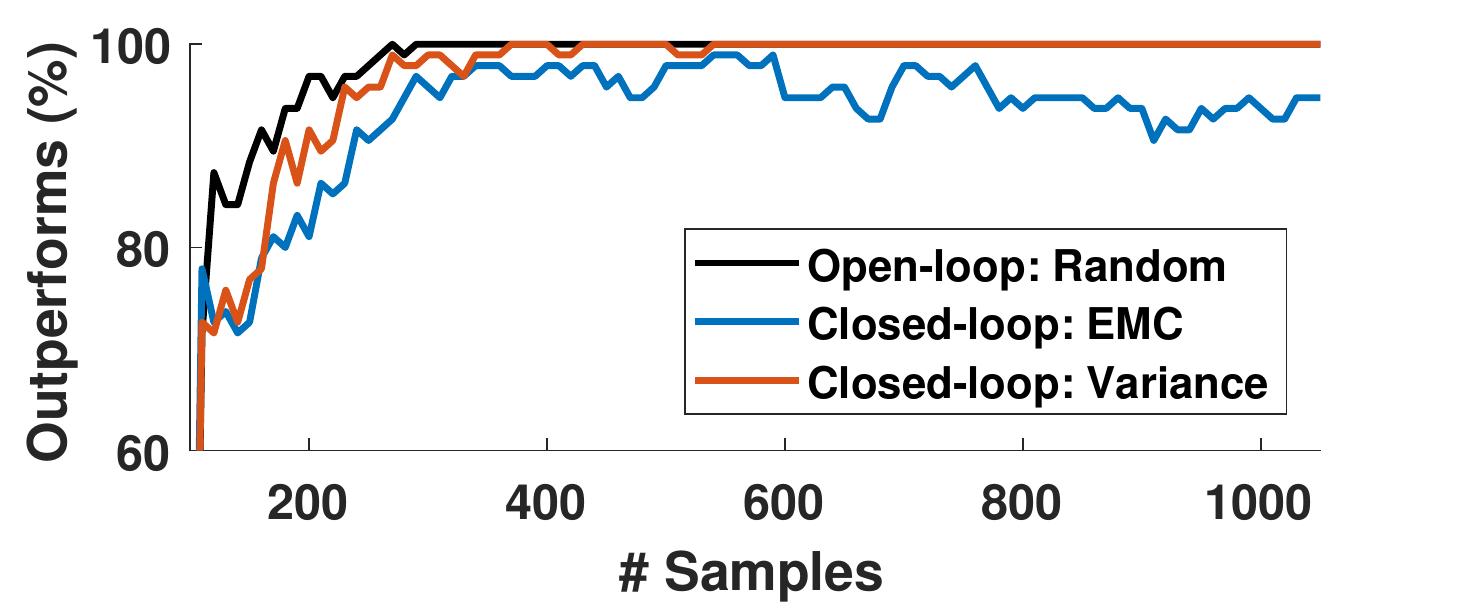}}
	\caption{(Example 2) Percentage of the 100 runs where the entropy-based procedure directly matches or outperforms the indicated approach given the same initial training dataset and GP model.}
 	\label{f:mrac2b}
	\vspace{-0.1in}
\end{figure}

Figure \ref{f:mrac2a} displays the total misclassification error for all four approaches while \fig{f:mrac2b} depicts the ratio of tests where Algorithm \ref{alg:entropyRMT} matches or outperforms the other three strategies.  The results reaffirm the conclusions from the previous example: Algorithm \ref{alg:entropyRMT} ultimately outperforms the other approaches in at least 95\% of tests, with a 33\% and 35\% improvement in mean misclassification error over the EMC- and variance-based procedures.  

\begin{figure}[!]
	\centering
	{\includegraphics[width=.79\columnwidth]{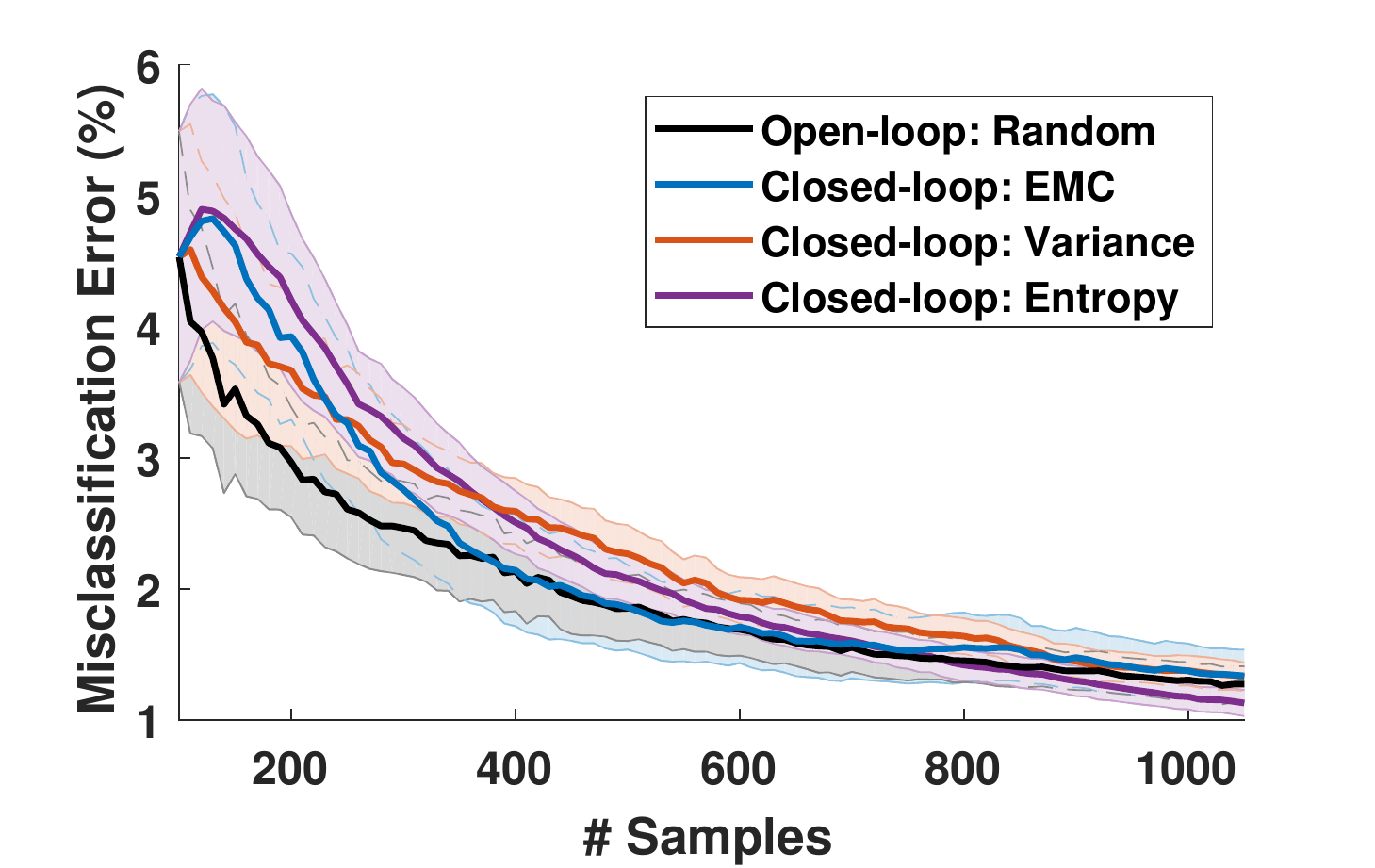}}
	\caption{(Example 2) Misclassification error within the 95\% prediction confidence levels. Unlike the total misclassification error, this only counts prediction errors for points with high confidence. Standard deviation intervals correspond to 0.5$\sigma$ bounds.}
 		\label{f:mrac2c}
	\vspace{-0.1in}
\end{figure}

Figure \ref{f:mrac2c} highlights the use of the prediction confidence \cref{eq:probSat} as an online validation tool to identify likely misclassification errors without external validation.  Each $\vtheta$ queried by the GP prediction model will return not just the binary prediction ($\vtheta \in \hat{\Theta}_{sat}$ or $\hat{\Theta}_{fail}$), but also a confidence level in that prediction.  Not surprisingly, the GP model is more likely to misclassify points with low confidence.  Figure \ref{f:mrac2c} displays the misclassification error rate for points with high $(\geq 95\%)$ confidence.  In comparison to \fig{f:mrac2a}, the rate of misclassification error is significantly lower, demonstrating the GP prediction model correctly identified regions where misclassifications are likely to occur.  These results also show that the prediction error is consistent across the various open- and closed-loop verification procedures.  This last fact confirms the utility of GP-based, data-driven verification and its prediction confidence value, regardless of the exact sampling strategy.

%%%%%%%%%%%%%%%%
\subsection{Example 3: Lateral-Direction Autopilot}
The third example examines an aircraft autopilot for controlling lateral-directional flight modes.  In particular, the ``heading-hold'' autopilot turns the aircraft to and holds a desired reference heading.  The simulation model of the autopilot and DeHavilland Beaver airframe is provided by the ``Aerospace Blockset'' toolbox in Matlab/Simulink\cite{Elliott16_S5}.  This simulation model includes numerous nonlinearities such as the full nonlinear aircraft dynamics, nonlinear controllers, and actuator models with position and rate saturations.  The provided ``heading hold'' autopilot then must satisfy several performance requirements.  

The requirement chosen for this example is the ``altitude-hold'' requirement of the heading autopilot.  In addition to turning the aircraft to the desired heading angle, the autopilot must also ensure the aircraft's altitude remains within 35 feet of the initial altitude when the reference command was given\cite{Elliott16_S5}.  This requirement can be written as 
\begin{equation}
	\varphi_{height} = \Box_{[0,T_f]} (35 - |x[t]- x[0]| \geq 0),
\end{equation}
where $T_f = 50$ seconds is the final simulation time and $x[t]$ is the aircraft altitude (in feet) at time $t$.  The performance metric is similar to the last example with
\begin{equation}
	y(\theta) = \underset{t' \in [0,T_f]}{\text{min }} \rho^{\varphi_{height}}[t'](\theta).
\end{equation}
The verification procedures test the satisfaction of $\varphi_{height}$ against different conditions of the initial Euler angles and longitudinal inertia, $\vtheta = $[roll(0), pitch(0), heading(0), $ I_{yy}$]$^T$.  While satisfaction of other autopilot requirements can be explored, such as heading angle overshoot or steady-state tracking error, the altitude-hold requirement dominated the other requirements during an initial trade-space exploration.  The space of allowable perturbations $\Theta$ spans roll(0): $[-60^{\degree},60^{\degree}]$, pitch(0): $[4^{\degree},19^{\degree}]$, heading(0): $[75^{\degree},145^{\degree}]$, and inertia $I_{yy}: [5430,8430] \ (kg\cdot m^2)$.  The desired reference heading was kept constant at $112^{\degree}$. Set $\Theta_d$ discretizes $\Theta$ into a normalized 4D grid with 937,692 possible locations.  

\begin{figure}[!]
	\centering
	{\includegraphics[width=.79\columnwidth]{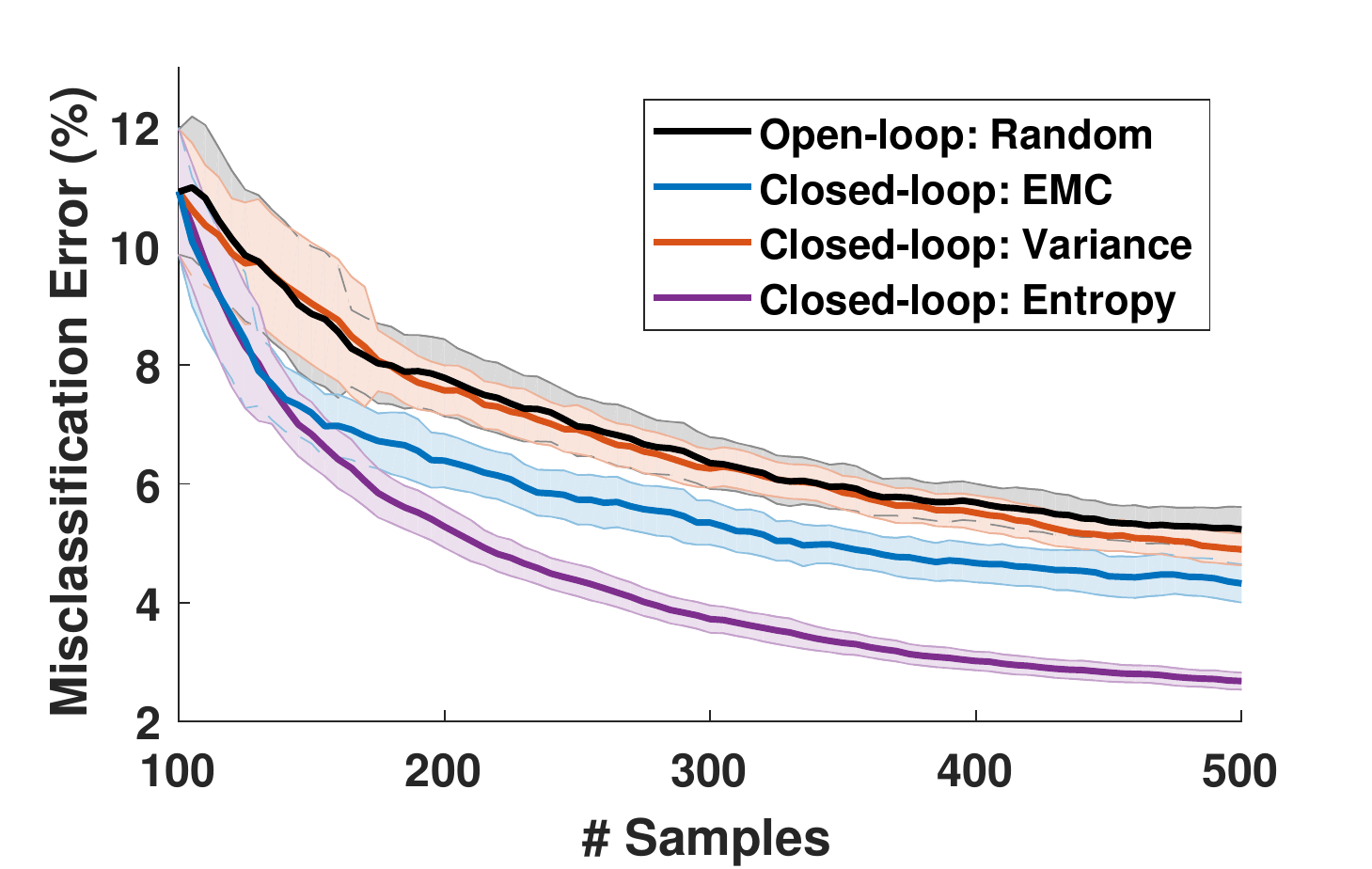}}
	\caption{(Example 3) Total misclassification error at batch size $M=5$ over 100 randomly-initialized runs.  Standard deviation intervals correspond to 0.5$\sigma$ bounds.}
 		\label{f:LMa}
	\vspace{-0.1in}
\vspace{0.2in}
	\centering
	{\includegraphics[width=.79\columnwidth]{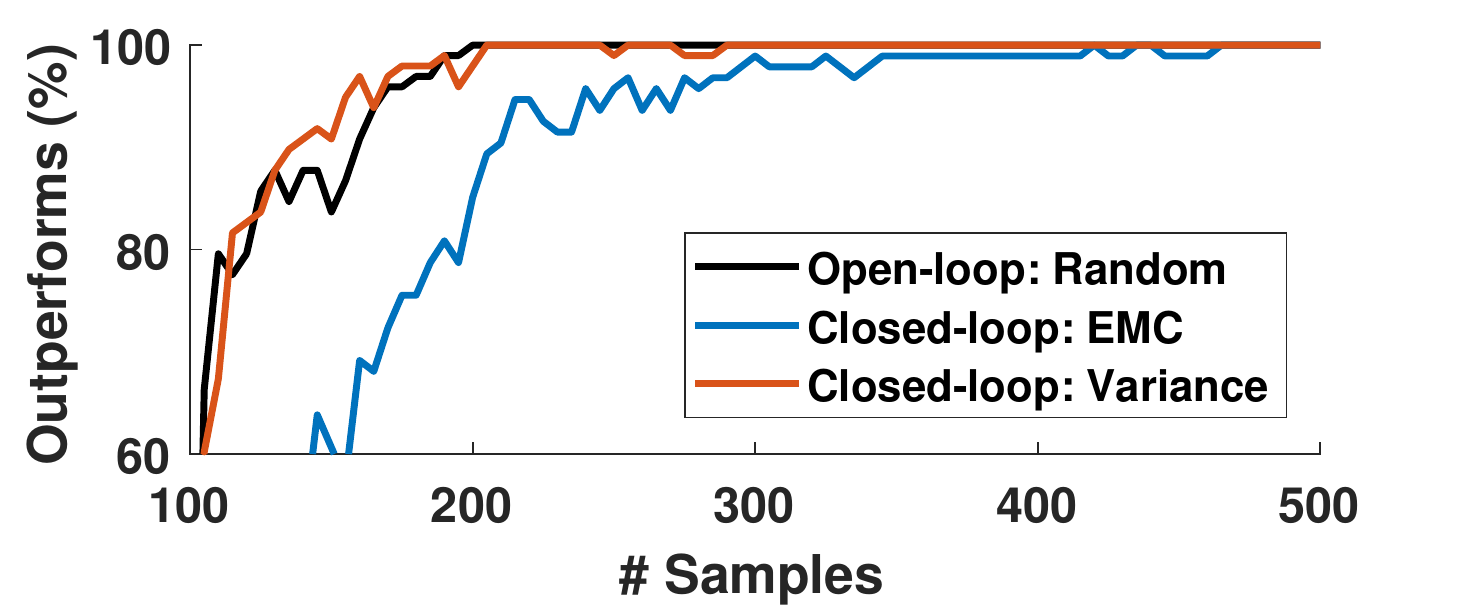}}
	\caption{(Example 3) Percentage of the 100 runs where the entropy-based procedure directly matches or outperforms the indicated approach given the same initial training dataset and GP model.}
 		\label{f:LMb}
	\vspace{-0.1in}
\end{figure}

This example uses the same general procedure from the last two examples.  During each of the 100 runs, the four procedures start with the same randomly-generated initial training set of 100 samples.  The closed-loop procedures then select 400 additional training samples in batches of $M=5$ points.  As seen in Figure \ref{f:LMa}, the entropy-based procedure outperforms the competing approaches to even greater extent than preceding examples.  The entropy-based procedure produces below $3\%$ average prediction error after 500 iterations, a $37\%$ improvement over the closest variance-based method and up to 50\% over the other two.  Similarly, the results in \fig{f:LMb} show that entropy-based, closed-loop verification will ultimately produce a lower prediction error than the competing algorithms 100\% of the time.  These results further highlight the consistently desirable performance of Algorithm \ref{alg:entropyRMT} for deterministic verification of complex nonlinear systems.

\begin{figure}[!]
	\centering
	{\includegraphics[width=.79\columnwidth]{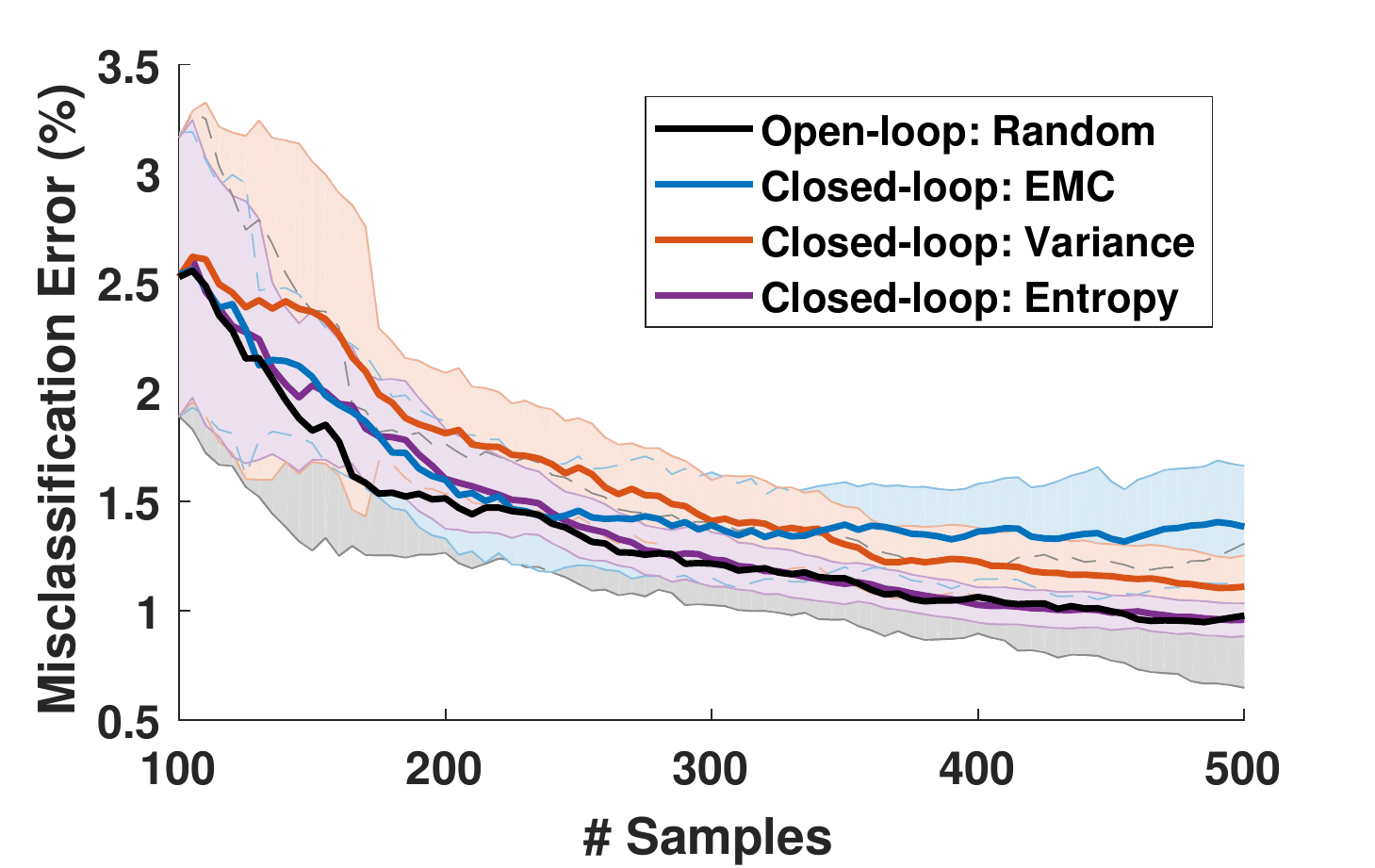}}
	\caption{(Example 3) Misclassification error within the 95\% prediction confidence levels. Unlike the total misclassification error, this only counts prediction errors for points with high confidence. Standard deviation intervals correspond to 0.5$\sigma$ bounds.}
 		\label{f:LMc}
	\vspace{-0.1in}
\end{figure}

Just like \fig{f:mrac2c}, \fig{f:LMc} demonstrates the ability of prediction confidence \cref{eq:probSat} to correctly identify points with low accuracy.  Once these points are removed, the misclassification error rate drops considerably, regardless of the sampling strategy.  Where as Figures \ref{f:LMa} and \ref{f:LMb} highlight the desirable impact of closed-loop statistical verification, \fig{f:LMc} reaffirms the utility of the online validation aspect of the new GP-based statistical verification framework.

%Additionally, this example more closely compares Algorithm \ref{alg:entropyRMT} against the other procedures.  Due to the fact all four procedures start from the same initial training set in each of the 100 runs, this work compares the four resulting trajectories directly with each other.  More specifically, Figure \ref{f:LMb} displays the percentage of runs where Algorithm \ref{alg:entropyRMT} directly outperformed the indicated approach.  The results show that entropy-based, closed-loop verification will ultimately produce a lower prediction error than the competing algorithms nearly 100\% of the time.

